\documentclass[10pt,aps,prd,nofootinbib,noshowkeys,noshowpacs,longbibliography,superscriptaddress,tightenlines,floatfix,twocolumn]{revtex4-2}
\usepackage{amsmath,amsfonts,amsthm,amssymb}
\usepackage{graphics,graphicx}
\usepackage{color}
\definecolor{darkblue}{RGB}{0,0,196}
\definecolor{darkgreen}{RGB}{0,120,0}

\def\HP{\hphantom{\alpha}} 

\usepackage{bm}

\def\beq{\begin{equation}}
\def\eeq{\end{equation}}
\def\ba{\begin{eqnarray}}
\def\ea{\end{eqnarray}}

\usepackage[colorlinks=true,linktocpage=true,linkcolor=darkblue,citecolor=red,urlcolor=darkblue]{hyperref}

\begin{document}
\preprint{}
 
    \title{Guiding center hydrodynamics with spin}
    \author{Rajeev Singh}
    \email{rajeev.singh@e-uvt.ro}
    \affiliation{Department of Physics, West University of Timisoara, Bulevardul~Vasile P\^arvan 4, Timisoara 300223, Romania}
	\date{\today} 
	\bigskip
\begin{abstract}
We build upon the recently formulated guiding-center kinetic theory for guiding-center plasma by incorporating spin degrees of freedom in the presence of electromagnetic fields. This approach yields a streamlined set of equations for guiding-center ideal hydrodynamics with spin--fewer than those in traditional spin hydrodynamics--owing to a restriction on motion perpendicular to the magnetic field. We propose that this framework offers a promising tool to comprehending how matter behaves under the influence of intense magnetic fields, such as spin polarization effects observed in experimental measurements.
\end{abstract}
     
\date{\today}
	
\maketitle
\newpage
\section{Introduction}
\label{sec:intro}
Relativistic hydrodynamic models have been employed to capture the collective behavior of various physical systems, such as cosmological and astrophysical plasmas, as well as the strongly-interacting plasma generated in heavy-ion collision experiments, among others~\cite{Bhattacharyya:2008ji,Bhattacharyya:2008xc,Haehl:2015foa,Romatschke:2017ejr,Andersson:2020phh}. These models have been thoroughly investigated to depict the bulk evolution of the strongly-interacting medium formed during relativistic heavy-ion collisions, proving effective in accounting for a range of collective phenomena. Both ideal and dissipative approaches have been formulated to interpret the hadron spectra observed in experiments. Given its achievements, it’s reasonable to assert that the hydrodynamic framework, combined with the characterization of the initial state and the method for hadronization, forms what can be considered the ‘standard model’ of heavy-ion collision physics~\cite{Hirano:2005wx,Teaney:2009qa,Heinz:2013th,Sorensen:2023zkk}.

The discovery of spin polarization in hyperons~\cite{STAR:2017ckg,STAR:2019erd,ALICE:2019aid} and spin alignment in vector mesons has unveiled exciting new avenues for exploring the characteristics of the hot, dense matter created in heavy-ion collisions. Yet, even as experimental data grows and future measurements are planned, the theoretical community has yet to reach agreement on the underlying causes of the observed spin polarization and alignment effects. The prevailing approach stems from quantum kinetic theory, offering a model where polarization arises purely from gradients of conventional hydrodynamic variables at the freeze-out stage. Interestingly, two distinct versions of this model successfully account for the Lambda polarization measurements in the top RHIC energy~\cite{Becattini:2021iol,Fu:2021pok}, which leaves lingering questions about the precise physical mechanism driving these fascinating polarization phenomena~\cite{Becattini:2024uha,Giacalone:2025bgm}.

In the light of these questions, an alternative viewpoint is to take into account the spin degrees of freedom in hydrodynamics. There has been intense investigations on `spin hydrodynamics' using various physical concepts~\cite{Becattini:2017gcx,Florkowski:2017ruc,Florkowski:2018ahw,Hattori:2019lfp,Fukushima:2020ucl,Li:2020eon,Weickgenannt:2020aaf,Shi:2020htn,Montenegro:2020paq,Bhadury:2020cop,Singh:2020rht,Gallegos:2021bzp,Weickgenannt:2021cuo,Cartwright:2021qpp,Florkowski:2021wvk,Hongo:2021ona,Yi:2021unq,Torrieri:2022ogj,Weickgenannt:2022zxs,Ambrus:2022yzz,Gallegos:2022jow,Singh:2022uyy,Daher:2022xon,Weickgenannt:2023bss,Biswas:2023qsw,Kumar:2023ojl,Jaiswal:2024urq,Bhadury:2024ckc,Wagner:2024fry,Singh:2024cub,Ren:2024pur,Florkowski:2024bfw,Drogosz:2024gzv,Chiarini:2024cuv,Lv:2024uev,Yang:2024duc,Daher:2025pfq,Bhadury:2025fil,Abboud:2025qtg}. The key object in spin hydrodynamic frameworks~\cite{Florkowski:2017ruc,Florkowski:2018ahw,Florkowski:2019qdp,Bhadury:2020cop} is the second-rank antisymmetric spin polarization tensor, $\omega^{\mu\nu}$, that characterizes the spin degrees of freedom.

Given the intense electromagnetic (EM) fields present in relativistic heavy-ion collisions, it becomes essential to incorporate these fields into spin hydrodynamics, as they may interact with spin degrees of freedom and influence the system’s evolution. How exactly EM fields couple with spin and shape its dynamics remains an intriguing, unresolved puzzle. Despite significant efforts in studies like~\cite{Singh:2021man,Bhadury:2022ulr,Peng:2022cya,Singh:2022ltu,Yang:2023zqe,Kiamari:2023fbe,Fang:2024sym}, which explore the complex interplay between spin and EM fields, the underlying physical mechanism continues to elude us.

A plasma of particles immersed in a strong magnetic field can be envisioned as a collection of guiding centers, and unraveling the collective dynamics of these guiding centers is key to comprehending how matter behaves under the influence of intense magnetic fields~\cite{Beklemishev736,Ripperda:2017ifb,Bacchini:2020cze,Mignone_2023,Trent:2023tnu,Trent:2024pii}. To effectively substitute the motion of particles with that of their guiding centers, the cyclotron radius should be smaller than the mean free path. In this work, we draw on the covariant guiding-center kinetic theory for guiding-center plasma, recently introduced in~\cite{Son:2024fgn}, and enhance it by integrating spin degrees of freedom using single particle spin dependent distribution function~\cite{Florkowski:2018fap} leading to the understanding of the effects of spin. This refined guiding-center hydrodynamics with spin results in a simpler framework, with fewer equations and reduced numerical complexity, offering a fresh perspective on these phenomena.

The article is organized as follows\footnote{We work with the metric signature $(+---)$ and the Levi-Civita symbol is defined as $\epsilon^{0123}=1$.}. We begin by revisiting the definition of the single-particle distribution function in Section~\ref{sec:distribution_function}, followed by an exploration of the conservation laws governing charge, the energy-momentum tensor, and the spin tensor in Section~\ref{sec:guiding_hydro}. The derivation of the constitutive relations is outlined in Section~\ref{sec:constitutive_relations}, while the entropy current is derived in Section~\ref{sec:entropy_current}. We summarize and provide outlook in Section~\ref{sec:concl}.

\section{Single particle spin dependent distribution function}
\label{sec:distribution_function}
We begin with the classical description of spin-$\frac{1}{2}$ particles with mass $m$ and intrinsic angular momentum $s^{\alpha\beta}$~\cite{Mathisson:1937zz,Florkowski:2018fap}
\begin{equation}
    s^{\alpha\beta}= \frac{1}{m}\epsilon^{\alpha\beta\gamma\delta}p_\gamma s_\delta\,,
    \label{eq:salbe}
\end{equation}
where $p_\gamma$ is the four-momentum of the particle and $s_\delta$ is the spin four-vector and in the rest frame of the particle read $p^\mu=(m,0,0,0)$ and $s^\mu=(0,\bm{s})$.
It is easy to notice that $s^{\alpha\beta}$ is antisymmetric in its indices and is also orthogonal to four-momentum, i.e. $p_\alpha s^{\alpha\beta}=0$, known as Frenkel condition.

The determination of the collisional invariants of the Boltzmann equation enables the construction of the equilibrium distribution functions~\cite{Florkowski:2018fap}
\begin{equation}
f^\pm_{\rm eq} (x,p,s) =  \exp\left(-\beta_\mu(x) p^\mu \pm \xi(x)+\frac{1}{2} \omega_{\mu\nu} (x) s^{\mu\nu}\right)\,,
\label{eq:feqxps}
\end{equation}
where $f^+_{\rm eq}$ and $f^-_{\rm eq}$ represents particle and antiparticle, respectively, with $\beta^\mu=\beta u^\mu=\frac{1}{T}u^\mu$ as the four-temperature and $\xi=\beta\mu$ is the ratio of baryon chemical potential and temperature.

In Eq.~\eqref{eq:feqxps}, $\omega_{\mu\nu}$ is the antisymmetric spin polarization tensor and can be seen as a conjugate to the spin angular momentum. Note that, since $s^{\mu\nu}$ is dimensionless, $\omega_{\mu\nu}$ is also dimensionless. In this work, we limit ourselves, for simplicity, to linear order in $\omega_{\mu\nu}$, hence, Eq.~\eqref{eq:feqxps} becomes\footnote{The dependence on $x,p,s$ and subscript `${\rm eq}$' are suppressed to avoid clutter.}
\begin{equation}
f^\pm =  \exp\left(-\beta_\mu \, p^\mu \pm \xi\right)\left(1+\frac{1}{2} \omega_{\mu\nu}  s^{\mu\nu} \right)\,.
\label{eq:feqxps1}
\end{equation}
One can integrate the spin degrees of freedom present in the distribution function to obtain the momentum-phase-space distribution function with the help of the spin integration measure~\cite{Florkowski:2018fap}
\begin{eqnarray}
    \int dS = \frac{2\,m}{\sqrt{3}\,\pi} \int d^4s   \,\delta\left(s\cdot s + \frac{3}{4}\right) \delta(p\cdot s)\,,
\end{eqnarray}
where the two delta functions represent the normalization of $s$ and the orthogonality conditions between $p$ and $s$. The coefficient outside the integral ensures the normalization
\begin{eqnarray}
    \frac{2\,m}{\sqrt{3}\,\pi} \int d^4s   \,\delta\left(s\cdot s + \frac{3}{4}\right) \delta(p\cdot s)=2\,,
\end{eqnarray}
indicating the spin degeneracy.

\section{Guiding-center hydrodynamics for spin polarized fluids}
\label{sec:guiding_hydro}
For a system comprising particles and antiparticles where spin degrees of freedom are accounted for solely via degeneracy factors, the key conserved quantities are the energy-momentum tensor ($T^{\mu\nu}$) and the charge current ($N^\mu$). When spin is explicitly taken into account, an additional conserved quantity emerges: the angular-momentum tensor ($J^{\lambda, \mu\nu}$)~\cite{Florkowski:2018fap,Florkowski:2018ahw}. This arises because the conservation law for total angular momentum in systems with spin exhibits a complex, non-trivial structure.

The total angular-momentum tensor ($J^{\lambda, \mu\nu}$) can be expressed as the combination of its orbital component ($L^{\lambda, \mu\nu}$) and its spin component ($S^{\lambda, \mu\nu}$), the latter being referred to as the spin tensor. It is widely recognized that multiple equivalent formulations of the energy-momentum and spin tensors exist, each capable of describing the system’s dynamics~\cite{Hehl:1976vr,Speranza:2020ilk,Singh:2024qvg}. The kinetic definitions adopted here align with those established by de Groot, van Leeuwen, and van Weert in \cite{DeGroot:1980dk}.

The forms of $T^{\mu\nu}$, $N^\mu$, and $S^{\lambda, \mu\nu}$ are linked to the properties of the system’s microscopic constituents via the moments of $f_{\rm eq}(x,p,s)$. By making use of the equilibrium distribution functions $f_{\rm eq}(x,p,s)$ as defined earlier, hydrodynamic quantities—such as the charge current, stress-energy tensor, and spin tensor—can be derived in a manner analogous to that of conventional hydrodynamics.

In this work, we employ the guiding-center kinetic theory~\cite{Son:2024fgn}. To simplify our analysis, we consider the existence of a potent electromagnetic field $F_{\mu\nu}$, characterized by magnetic dominance, where $F_{\mu\nu}F^{\mu\nu} = 2(\bm{B}^2 - \bm{E}^2) > 0$. At any point in spacetime, there is a reference frame in which the electric field $\bm{E}$ aligns parallel to the magnetic field $\bm{B}$. In this frame, we define $B_*$ and $E_*$ as the magnitudes of the magnetic and electric fields, respectively, with $E_* < 0$ when the fields are antiparallel. Expressed through Lorentz invariants, this becomes $F_{\mu\nu}F^{\mu\nu} = 2(B_*^2 - E_*^2)$ and $F_{\mu\nu}\tilde{F}^{\mu\nu} = -4E_*B_*$. We also assume that $B_* \gg E_*$.

We focus on a plasma composed of a single species of charged particles, with a stationary background charge density ensuring overall neutrality. While the electromagnetic field is generally considered static, it can be allowed to evolve by incorporating Maxwell's equations into the hydrodynamic framework.

We have the conservation laws
\begin{eqnarray}
    \partial_\mu N^\mu &=& 0\,,
    \label{eq:current_conservation}\\
    \partial_\mu T^{\mu\nu} &=& F^{\nu\mu}J_\mu\,,
    \label{eq:EMT_conservation}
\end{eqnarray}
where the charge current $N^\mu=N^{\mu}_\parallel+N^{\mu}_\perp$ is decomposed into longitudinal current, $N^{\mu}_\parallel$ (parallel to the magnetic field lines), and transverse current, $N^\mu_\perp$ (perpendicular to the magnetic field lines), also known as drift current~\cite{Son:2024fgn}. 

Similarly, $T^{\mu\nu}$ is decomposed into parallel and perpendicular parts, $T^{\mu\nu}=\tilde{T}^{\mu\nu}+T^{\mu\nu}_\perp$ where $T^{\mu\nu}_\perp$ is perpendicular to the magnetic field arising from the particle's transverse motion in the presence of the magnetic field. Here, $J^\mu=N^\mu + M^{\mu}$ is the sum of charge current and magnetization current. The magnetization current reads
\begin{equation}
    M^\mu = \partial_\sigma M^{\sigma\mu}\,,
\end{equation}
where $M^{\sigma\mu} = ({M}/{B_*}) B^{\sigma\mu}$ is the magnetization density coming from the particle's transverse motion with $M$ being the magnetic moment density. As $T^{\mu\nu}$ is symmetric, we will have an additional conservation equation for the spin tensor
\begin{equation}
    \partial_\lambda S^{\lambda,\mu\nu} = 0\,.
    \label{eq:spin_tensor}
\end{equation}
We shall now derive the constitutive relations for $N^\mu$, $T^{\mu\nu}$, and $S^{\lambda,\mu\nu}$ along with the form of $M$ using the spin dependent distribution function Eq.~\eqref{eq:feqxps1}.
\section{Constitutive relations}
\label{sec:constitutive_relations}
We consider the physical situation as presented in the introduction where magnetic field dominates over the electric field $(B_*\gg E_*)$ and in the frame where $\bm{E}\parallel \bm{B}$. Although the guiding center’s momentum is consistently aligned with the magnetic field owing to the restriction $B^{\mu\nu}p_\nu = 0$, the particle’s motion isn’t confined solely to the magnetic field lines. It also exhibits a gentle drift perpendicular to the field, with a velocity described by the spatial components of $v^\mu$, also known as drift velocity~\cite{Son:2024fgn}.
\subsection{Charge current}
\label{subsec:charge_current}
Considering the physical scenario described above, the charge current is defined as
\begin{eqnarray}
    N^\mu &=& \int dP\, dS\, \left(p^\mu+v^\mu\right) \left(f^+ -f^- \right)\, ,\nonumber\\
    &=& N^\mu_\parallel + N^\mu_\perp\,,
    \label{eq:Nmu_relation}
\end{eqnarray}
where~\cite{Son:2024fgn} 
$$v^\mu = B_*^{-2}\left(p_\nu p^\lambda  \partial_\lambda B^{\mu\nu} + \frac{\tilde m^2 -m^2}{2 B^*}B^{\mu\nu}\partial_\nu B_*\right),$$
represents the slow drift velocity of the guiding center due to field inhomogeneities and curvature, not the cyclotron motion itself. It arises from the first-order correction to the guiding-center momentum equation. Hence, $v^{\mu}$ encodes the averaged $(E\times B)$ and curvature drifts that are perpendicular to $B^{\mu\nu}$.

The momentum integration measure is defined as~\cite{Son:2024fgn}
\begin{equation}
    dP = \frac{d^4 p}{2\pi^2} \theta(p_0)\,\delta(p^2-\tilde{m}^2)\,\delta^2(\Delta^\mu_{\HP\nu}p^\nu)\left(B_*+\frac{p_\mu \partial_\nu B^{\mu\nu}}{B_*}\right),
    \label{eq:dP}
\end{equation}
in which the overall factors are chosen to be consistent with Landau level degeneracy. Note that, $\theta(p_0)\,\delta(p^2-\tilde{m}^2)$ takes care of the mass-shell condition and $\delta^2(\Delta^\mu_{\HP\nu}p^\nu)$ forces the phase space to have only one momentum direction, $\bm{p}\parallel \bm{B}$. Here, $p_0^2=p_\parallel^2+\tilde m^2$ with $\tilde m^2 = \overline{p_\perp^2} + m^2$, and $\Delta^\mu_{\HP\nu}=(1/B_*^2)B^{\mu\alpha}B_{\nu\alpha}$ is the projector onto the transverse plane $(x,y)$ perpendicular to the $\bm{B}$ plane in the $\bm{E}\parallel \bm{B}$ frame. Note that, $N^{\mu}_\parallel$ and $N^\mu_\perp$ satisfy $B_{\mu\nu}N^\nu_\parallel=0$ and $\tilde{\Delta}^\mu_\nu N^\nu_\perp=0$, respectively. Here, $B_{\mu\nu}$ is the magnetic part of $F_{\mu\nu}$ and $\tilde\Delta^\mu_{\HP\nu} = \delta^\mu_\nu - \Delta^\mu_{\HP\nu}$ is the projector onto the 1+1 $\bm{E}\parallel \bm{B}$ space, i.e., $(t,z)$ hyperplane.

To obtain the longitudinal current, we first put Eq.~\eqref{eq:feqxps1} in the first part of Eq.~\eqref{eq:Nmu_relation}
\begin{eqnarray}
    N_\parallel^\mu &=& 2\,\sinh{(\xi)}\int dP\, dS\, p^\mu e^{-\beta \cdot p}\left(1+\frac{1}{2} \omega_{\mu\nu}  s^{\mu\nu} \right),\,\,\,
\end{eqnarray}
where the spin integration gives
\begin{eqnarray}
    \int dS \left(1+\frac{1}{2} \omega_{\mu\nu}  s^{\mu\nu} \right)&=& 2\,,
\end{eqnarray}
as $\int dS s^{\mu\nu}=0$~\cite{Florkowski:2018fap}. Thus, we receive,
\begin{eqnarray}
     N_\parallel^\mu &=& 4\,\sinh{(\xi)}\int dP\,  p^\mu e^{-\beta \cdot p}\,.
\end{eqnarray}
The above integral can be computed in the rest frame of the fluid, $u=\left(1,0,0,0\right)$, and then applying a boost in the magnetic field direction which results in
\begin{eqnarray}
     N_\parallel^\mu &=& n\,u^\mu = 4\,\sinh{(\xi)}\,n_0\, u^\mu\,,
\end{eqnarray}
where 
\begin{eqnarray}
    n_{0} &=& \int dP \, (u\cdot p)\,  e^{- \beta \cdot p} = \frac{B_*}{(2 \pi)^2} \, \tilde{m} \, K_1 (\tilde{z})\,.
\end{eqnarray}
Here, $K_1$ is the modified Bessel function of the second kind and $\tilde{z}=\tilde{m}/T$. We used the following relations to compute the thermal integrals in this work
\begin{eqnarray}
    \int^\infty_z d\tau \left(\tau^2-z^2\right)^{n-\frac{1}{2}}e^{-\tau}&=& z^n \frac{(2n)!}{2^n \, n!}K_n(z)\,,\\
    \int^\infty_z d\tau \left(\tau^2-z^2\right)^{n-\frac{3}{2}}\tau\,e^{-\tau}&=& z^n \frac{(2n-2)!}{2^{n-1} \,(n-1)!}K_n(z)\,.\nonumber
\end{eqnarray}
Similarly, after substituting $v^\mu$ in Eq.~\eqref{eq:Nmu_relation} we obtain
\begin{eqnarray}
    N^\mu_\perp &=& B_*^{-2}\left(\tilde{T}^\lambda_{\HP\nu}  \partial_\lambda B^{\mu\nu} - M B^{\mu\nu}\partial_\nu B_*\right)\,.
    \label{eq:Nmu_perp}
\end{eqnarray}
The forms for $\tilde T^\lambda_{\HP\nu}$ and $M$ will be derived in the next section. Due to the presence of derivatives in \eqref{eq:Nmu_perp}, $N_\perp^\mu \ll N_\parallel^\mu$, hence, can be neglected for ideal spin hydrodynamics~\cite{Son:2024fgn}.

\subsection{Energy-momentum tensor}
\label{eq:subsec:EMT}
The definition of the energy-momentum tensor reads
\begin{eqnarray}
    T^{\mu\nu} &=& \int dP \,dS\, \left(p^\mu p^\nu+ \overline{p^\mu_\perp p^\nu_\perp}\right) \left(f^+ + f^- \right)\, ,\nonumber\\
    &=& \tilde{T}^{\mu\nu}+T^{\mu\nu}_\perp\,.
    \label{eq:Tmunu}
\end{eqnarray}
Note that, $\overline p_{\perp}^{\mu}\equiv\Delta^{\mu}_{\,\,\,\,\nu}p^{\nu}$ denotes the residual, gyromotion-averaged perpendicular momentum that appears after projecting out the fast cyclotron motion. It accounts for the finite magnetic-moment contribution that survives under the guiding-center approximation. Although the integration measure $(dP)$ contains $(\delta^{2}(\Delta^{\mu}_{\,\,\,\,\nu}p^{\nu}))$ to restrict the on-shell momentum to the magnetic-field direction, the transverse kinetic contribution survives because $(\overline{p_{\perp}^{2}})$ enters through the effective mass $(\tilde m^{2}=m^{2}+\overline{p_{\perp}^{2}})$. The $(\delta^{2})$ constraint removes explicit dependence on $(\overline p_{\perp}^{\mu})$ in phase space but not its averaged squared magnitude, which manifests as the transverse pressure $(P_{\perp}\propto(\tilde m^{2}-m^{2}))$, see Eq.~\eqref{eq:Moment}.

Therefore, using Eq.~\eqref{eq:feqxps1}, we have
\begin{eqnarray}
    \tilde{T}^{\mu\nu}&=& 4\, \cosh(\xi) \int dP\, p^\mu p^\nu e^{-\beta \cdot p}\, ,
\end{eqnarray}
where again we use the facts that $\int dS=2$ and $\int dS \, s^{\mu\nu}=0$. Thus,
\begin{eqnarray}
    \tilde T^{\mu\nu}= (\epsilon + P) u^\mu u^\nu - P\tilde\Delta^{\mu\nu}\,,
    \label{eq:TmunuTilde}
\end{eqnarray}
where
\begin{eqnarray}
    \epsilon &=& 4\,\cosh(\xi)\,\int dP \,(u\cdot p)^2 e^{-\beta \cdot p}= 4\,\cosh(\xi)\,\epsilon_0\,,
\end{eqnarray}
with $$\epsilon_0 = \frac{B_*}{(2\pi)^2}\tilde{m}\,T \left(\tilde z K_0(\tilde z)+K_1 (\tilde z)\right)$$
and 
\begin{eqnarray}
    P &=& 4\,\cosh(\xi)\int dP\, \frac{1}{3} \left((u\cdot p)^2 - p\cdot p \right) e^{-\beta \cdot p}\,,\nonumber\\
    &=& 4\,\cosh(\xi)\,P_0\,,
\end{eqnarray}
where $P_0 = \frac{B_*}{(2\pi)^2}\tilde m \, T\, K_1(\tilde z)$.
Note that, $P_0 = n_0 \, T$ which is the expected state equation of ideal relativistic Boltzmann statistics. In Eq.~\eqref{eq:TmunuTilde}, $\tilde\Delta^\mu_{\HP\nu} = \delta^\mu_\nu - \Delta^\mu_{\HP\nu}$ is the projector onto the 1+1 $\bm{E}\parallel \bm{B}$ space, i.e., $(t,z)$ hyperplane.

Similarly, the transverse part of the energy-momentum tensor in the magnetic field dominated system where $\bm{E}\parallel \bm{B}$ is defined as, using Eq.~\eqref{eq:feqxps1} in Eq.~\eqref{eq:Tmunu} and doing the spin integration,~\cite{Son:2024fgn},
\begin{eqnarray}
    T^{\mu\nu}_\perp &=& 4\, \cosh(\xi)\int dP \, \overline{p^\mu_\perp p^\nu_\perp}\, e^{-\beta \cdot p}\,,\nonumber\\
    &=& - 2\, \cosh(\xi)\int dP \, \Delta^{\mu\nu}p_\perp^2\, e^{-\beta \cdot p}\,,\nonumber\\
    &=&-P_\perp \Delta^{\mu\nu}\,,
\end{eqnarray}
where the transverse pressure (due to gyromotion) is
\begin{eqnarray}
    P_\perp &=& 2 \cosh(\xi) P_{\perp 0}\,,\nonumber\\
    &=&2 \cosh(\xi)\frac{B_*}{(2\pi)^2}(\tilde m^2 - m^2)K_0(\tilde z)\,,\nonumber\\
    &=&-M \, B_*\,.
    \label{eq:Moment}
\end{eqnarray}
Using Eqs.~\eqref{eq:Nmu_perp} and \eqref{eq:Moment} simplifies the conservation of energy-momentum tensor to
\begin{eqnarray}
    \tilde \Delta^\sigma_{\HP\nu}\partial_\mu T^\mu_{\HP\sigma}=E_{\nu\lambda} N_\parallel^\lambda\,,
\end{eqnarray}
where the placing of $\tilde\Delta^\sigma_{\HP\nu}$ indicates that there are two independent equations rather than four~\cite{Son:2024fgn}. 
$E_{\nu\lambda}$ is the electric part of the electromagnetic tensor corresponding to the electric field, where $E^{\mu}=F^{\mu\nu}u_{\nu}$ is the electric-field four-vector in the fluid frame.

It is important to highlight that, here $u^\mu$ is defined by the conditions, $u^\mu u_\mu=1$, $B_{\mu\nu}u^\nu = 0$, and $\tilde\Delta^\mu_{\HP\sigma}T^\sigma_{\HP\nu}u^\nu = \epsilon u^\mu$ which leave velocity with only one degree of freedom.

\subsection{Spin tensor}

The key ingredient in our formalism is the spin tensor which is defined as the moment of the intrinsic angular momentum~\cite{DeGroot:1980dk,Florkowski:2018fap}
\begin{eqnarray}
    S^{\lambda,\mu\nu} &=& \int dP\,dS\, p^\lambda\, s^{\mu\nu}\left(f^+ + f^- \right), \label{eq:spin}\\
    &=&2\,\cosh(\xi)\int dP\,dS\, p^\lambda s^{\mu\nu}e^{-\beta\cdot p}\left(1+\frac{1}{2} \omega_{\alpha\beta}  s^{\alpha\beta}\right).\nonumber
\end{eqnarray}
The spin integration gives
\begin{eqnarray}
    \int dS\, s^{\mu\nu}\left(1+\frac{1}{2} \omega_{\alpha\beta}  s^{\alpha\beta}\right) &=& \frac{\omega^{\mu\nu}}{2}+ \frac{p^\alpha}{\tilde m^2}p^{[\mu}\omega^{\nu]}_{\HP\alpha}\,.
    \label{eq:spin_integral}
\end{eqnarray}
The momentum integration is a bit tricky due to the measure defined in Eq.~\eqref{eq:dP} and needs a careful treatment. Using Eq.~\eqref{eq:spin_integral} in Eq.~\eqref{eq:spin}, we obtain
\begin{eqnarray}
    S^{\lambda,\mu\nu}&=& S^{\lambda,\mu\nu}_1 + S^{\lambda,\mu\nu}_2\,,
\end{eqnarray}
where
\begin{eqnarray}
    S^{\lambda,\mu\nu}_1 &=& \omega^{\mu\nu}\cosh(\xi) \int dP\, p^\lambda\, e^{-\beta \cdot p}
    = \frac{P}{4} \beta^\lambda \, \omega^{\mu\nu}\,,
    \label{eq:spin1}
\end{eqnarray}
and
\begin{eqnarray}
    S^{\lambda,\mu\nu}_2 &=& 2\,\cosh(\xi) \int dP\, p^\lambda\, e^{-\beta \cdot p}\frac{p^\alpha}{\tilde m^2}p^{[\mu}\omega^{\nu]}_{\HP\alpha}\,,\nonumber\\
    &=&T^2\left(\left(\epsilon+P\right)\frac{3\,T^2}{\tilde m^2}+\frac{P}{2}\right) \beta^\lambda \beta^\delta \beta^{[\mu} \omega^{\nu]}_{\HP\delta}\label{eq:spin2}\\
    &-&\frac{T^2}{2\tilde m^2}\left(\epsilon+P\right)\bigg\{\tilde \Delta^{\lambda[\mu}\omega^{\nu]}_{\HP\delta} \beta^\delta
    +\tilde \Delta^{\delta[\mu}\omega^{\nu]}_{\HP\delta} \beta^\lambda\nonumber\\
    &+& \tilde \Delta^{\lambda\delta}\beta^{[\mu}\omega^{\nu]}_{\HP\delta}\bigg\}.\nonumber
\end{eqnarray}
Thus, the spin tensor, Eq.~\eqref{eq:spin}, looks like
\begin{eqnarray}
    S^{\lambda,\mu\nu}&=& \frac{P}{4} \beta^\lambda \omega^{\mu\nu}+ T^2\left(\left(\epsilon+P\right)\frac{3\,T^2}{\tilde m^2}+\frac{P}{2}\right) \beta^\lambda \beta^\delta \beta^{[\mu} \omega^{\nu]}_{\HP\delta}\nonumber\\
    &-&\frac{T^2 \left(\epsilon+P\right)}{2\tilde m^2}\bigg\{\tilde \Delta^{\lambda[\mu}\omega^{\nu]}_{\HP\delta} \beta^\delta
    +\tilde \Delta^{\delta[\mu}\omega^{\nu]}_{\HP\delta} \beta^\lambda\nonumber\\
    &+&\tilde \Delta^{\lambda\delta}\beta^{[\mu}\omega^{\nu]}_{\HP\delta}\bigg\}.
\end{eqnarray}
Since we work with the linear-order in $\omega_{\mu\nu}$, there is no feedback from spin conservation to the conservation laws of charge current and energy-momentum tensor.

\section{Entropy current}
\label{sec:entropy_current}
Ideal spin hydrodynamics must conserve entropy. To show that, we start with the entropy current definition
\begin{eqnarray}
    \mathfrak{S}^\mu &=& -\int dP\,dS\, p^\mu \left(f^+ \log(f^+) + f^- \log(f^-)\right)\nonumber\\
    &&+\int dP\,dS\, p^\mu\left(f^+ + f^-\right)\,.
    \label{eq:Entropy_current}
\end{eqnarray}
Using Eq.~\eqref{eq:feqxps} and conservation laws for charge current, energy-momentum tensor, and spin tensor, we have
\begin{eqnarray}
    \mathfrak{S}^\mu &=& \beta_\alpha T^{\mu\alpha}-\frac{1}{2}\omega_{\alpha\beta} S^{\mu,\alpha\beta} + N^\mu \left(\coth(\xi)-\xi\right)\,.
\end{eqnarray}
Hence, the divergence of the entropy current results into
\begin{eqnarray}
    \partial_\mu \mathfrak{S}^\mu &=& T^{\mu\alpha} \partial_\mu \beta_\alpha + \beta_\alpha \left(\tilde{\Delta}^{\lambda\alpha}\partial_\mu \tilde{T}^\mu_{\HP\lambda}-E^\alpha_\lambda N_\parallel^\lambda\right)\nonumber\\
    &-& \frac{1}{2}S^{\mu,\alpha\beta}\partial_\mu \omega_{\alpha\beta}-N^\mu (\coth(\xi))^2 \partial_\mu \xi\,,
\end{eqnarray}
which, using Eqs.~\eqref{eq:feqxps}, \eqref{eq:Nmu_relation}, \eqref{eq:current_conservation} and requiring that terms involving $\partial \cdot u$ cancel out separately from those containing $u\cdot \partial$ derivatives, gives 
\begin{equation}
 \partial_\mu \mathfrak{S}^\mu=0   \,.
\end{equation}
In the above, considering the case of ideal (zeroth-order) spin hydrodynamics we neglected the contribution of $N^\mu_\perp$ to $N^\mu$ and restricted ourselves to leading-order in $\omega^{\mu\nu}$. The entropy conservation dictates that the thermodynamic relationship between the thermodynamic coefficients 
\begin{eqnarray}
    n&=&\frac{\partial P}{\partial \mu}\,,\nonumber\\ \epsilon&=&T\frac{\partial P}{\partial T}+ \mu\frac{\partial P}{\partial \mu}-P\,,\nonumber\\
    P_\perp &=& P - B_*\frac{\partial P}{\partial B_*}.
\end{eqnarray}
is satisfied, which is indeed the case for our spin hydrodynamic framework.

It’s worth highlighting that the contributions to entropy production arising from the spin polarization tensor are quadratic in nature. This implies that, at the linear order, spin polarization has no impact on entropy production. For both the conserved charge and the energy-momentum tensor, any corrections from $\omega_{\alpha\beta}$ begin at the second order~\cite{Florkowski:2024bfw}.

\section{Conclusion \& Outlook}
\label{sec:concl}
In this study, we have enriched the recently developed guiding-center ideal hydrodynamics by incorporating spin degrees of freedom. This enhancement enables the formulation of a kinetic theory and the derivation of ideal spin hydrodynamic equations for a fluid of charged particles immersed in a strong magnetic field background. This approach yields a fewer number of spin hydrodynamic equations for guiding-center ideal hydrodynamics with spin—than those in traditional spin hydrodynamics—owing to a restriction on motion perpendicular to the magnetic field. 

Looking ahead, there are several exciting avenues to expand this work and are ongoing. For instance, integrating dissipation and particle collisions would lend greater realism to the framework. Additionally, exploring beyond the linear order in $\omega_{\mu\nu}$ to investigate spin feedback effects promises to be a compelling pursuit. Lastly, while we have established the spin hydrodynamic relations within a static electromagnetic field, this could be evolved into a dynamic system by coupling it with Maxwell’s fields.
\section*{Acknowledgements}
R.S. acknowledges support from a postdoctoral fellowship at the West University of Timișoara, Romania. Valuable discussions with Victor Ambrus, Aritra Bandyopadhyay, Misha Stephanov, and Derek Teaney are gratefully acknowledged. We also thank Dirk Rischke for his critical reading of the manuscript.
\bibliographystyle{utphys}
\bibliography{pv_ref}
\end{document}